# Acoustic Structure Inverse Design and Optimization Using Deep Learning

Xuecong Sun[1,2], Yuzhen Yang[1*], Han Jia[2,3], Han Zhao[1,2], Yafeng Bi[1], Zhaoyong Sun[1] and Jun Yang[1,2*]

[1] Key Laboratory of Noise and Vibration Research, Institute of Acoustics, Chinese Academy of Sciences, Beijing 100190, People's Republic of China

[2] University of Chinese Academy of Sciences, Beijing 100049, People's Republic of China

[3] State Key Laboratory of Acoustics, Institute of Acoustics, Chinese Academy of Sciences, Beijing 100190, People's Republic of China

Correspondence should be addressed to Yuzhen Yang, yangyuzhen@mail.ioa.ac.cn, Jun Yang; jyang@mail.ioa.ac.cn

Keywords: acoustics structure design, deep learning, multi-order Helmholtz resonator, sound insulation

## Abstract

From ancient to modern times, acoustic structures have been employed to manage the spread of acoustic waves. Nevertheless, designing these structures traditionally remains a laborious and computationally intensive iterative process. Recognizing that complex acoustic systems can be effectively analyzed using the lumped-parameter method, we introduce a deep learning model that learns the correlation between the equivalent electrical parameters and the acoustic properties of these structures. As an illustration, we consider the design of multi-order Helmholtz resonators, showing experimentally that our model can predict structures with high precision that closely align with the specified design criteria. Furthermore, our model can seek multiple solutions in conjunction with dimensionality reduction algorithms and support evolutionary algorithms in optimization tasks. Compared to traditional numerical methods, our approach offers greater efficiency, flexibility, and universality. The designed acoustic structures hold broad potential for applications including speech enhancement, sound absorption, and insulation.

## 1. Introduction

Acoustic structures have been used for centuries to control acoustic waves in terms of amplitude and phase. With extensive achievements in fabrication technology, unprecedented functionalities can be obtained by designing and engineering artificial structures with more complex properties. Recently, many exotic functionalities, such as anomalous refraction/reflection [1,2], invisibility [3,4], timbre manipulation [5] and novel acoustic sensing [6,7], have been realized with fantastic acoustic structures. However, the development of accurate and computationally efficient design and optimization approaches for the acoustic structures is still in the early stages. During the design process, the forward calculation, i.e., predicting the acoustic properties based on the geometric structures, is well understood with analytical and numerical approaches, such as the lumped-parameter technique (LPT), transfer matrix method (TMM) and finite element method (FEM) and modal displacement method [8]. Nevertheless, the inverse problem, i.e., inferring acoustic structures from on-demand acoustic properties, is currently a prohibitive task even with the most advanced numerical tools. To search the formidably large design space efficiently, the inverse design procedure is usually guided by optimization algorithms, such as gradient-based approaches and evolutionary approaches. As structural complexity grows, the above methods will take a prohibitive amount of time, which seriously limits the usefulness of such approaches. Therefore, it is of great significance to identify an efficient, flexible and universal acoustic structure design method.

In recent years, deep learning has emerged as a very powerful computational method. In light of its exceptional success in domains related to computer science and engineering [9–13], deep learning has attracted increasing attention from researchers in other disciplines, including materials science [14], chemistry [15], physics [16–18], computational imaging and microscopy [19–21]. Moreover, deep learning has become a radically new approach in the context of photonic and electromagnetic design, such as approximation of light scattering from plasmonic nanostructures and inverse design of the electromagnetic metasurface structure, over the past few years [22–27]. Recently, deep learning has also been used to solve the inverse problem of the nonlocal metasurface, variable cross-sectional acoustic structure and two-dimensional acoustic cloaking [28–31]. However, both the researches trained the deep neural networks (DNNs) for specific structures, which are hard to be extended to other acoustic structures.

In view of the current problems and previously reported designs, here we propose an efficient, flexible and universal acoustic structure design method to solve the inverse problem. In our method, we analyze acoustic structures by LPT and develop connections between geometric parameters (GPs) and equivalent electrical parameters (EEPs). Then, a DNN is set up and trained to determine the corresponding relation between the EEPs of the acoustic structures and their acoustic properties. To evaluate the effectiveness of the proposed design method, a multi-order Helmholtz resonator is designed to realize acoustic insulation at specific frequencies (see Figure 1a). Once design requirements are input into the trained model, the EEPs of the structures will be generated quickly and automatically. Then, the GPs can be calculated through the LPT. In this example, it has been proven that the trained DNN can not only give a satisfied solution of the inverse design problems more quickly than its numerical counterpart, but also be used to search for multiple solutions combined with the dimension reduction algorithm. Moreover, it can also be used to improve the performance of evolutionary approaches in optimization for a desired property. Finally, we design an acoustic filter to decrease the line-spectrum noise using the proposed design method. This deep learning approach is an effective design tool for acoustic structure on-demand design and optimization. Considering that various acoustic structures can be analyzed by the LPT exactly in the low frequency range, such as Mie resonant structure [32], membrane-cavity acoustic absorber [33] and acoustic structure of microphone [34], the proposed approach has a strong versatility and scalability, which can be further extended to other acoustic structures.

## 2. Method

### 2.1 Physical Model

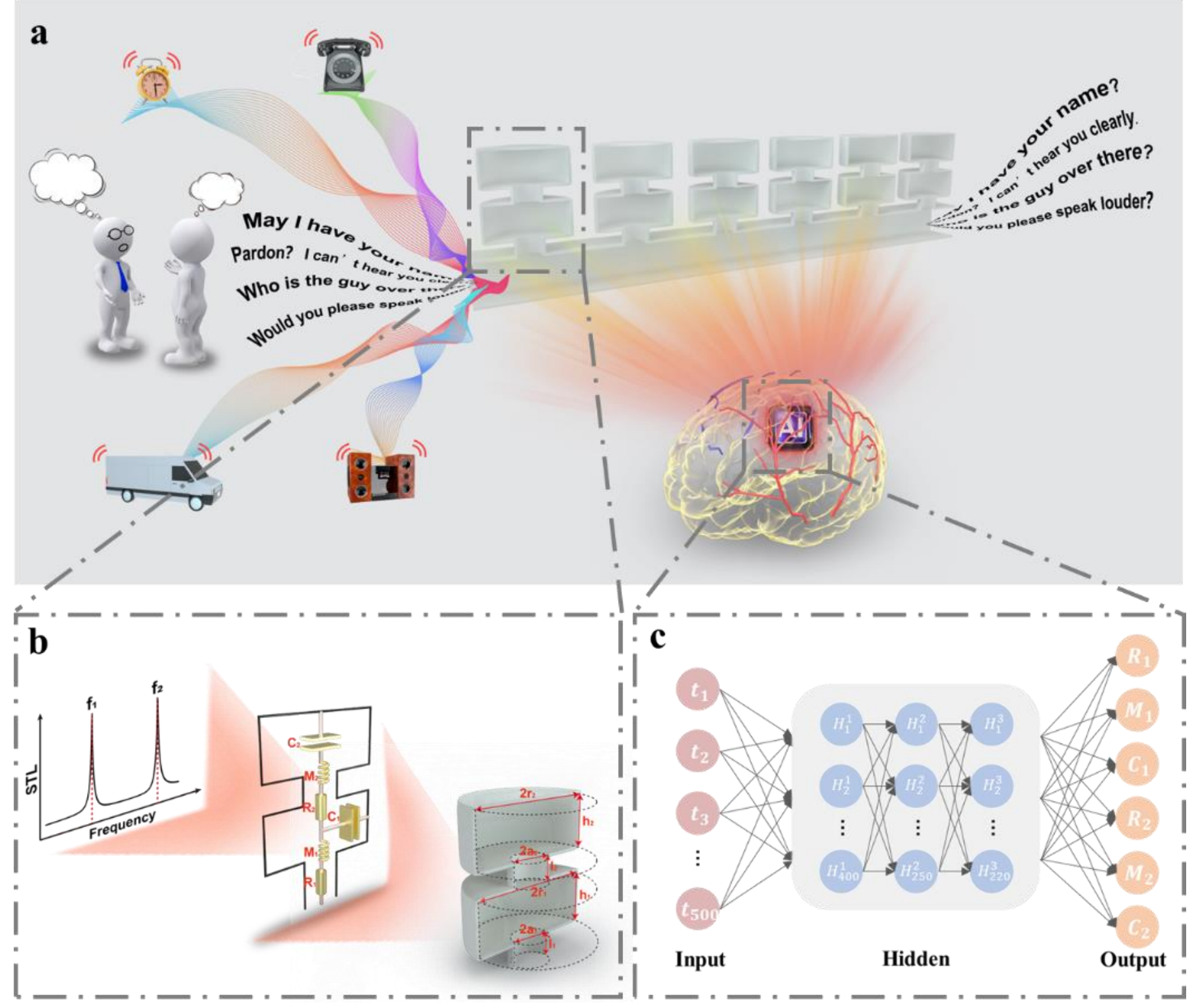

**Figure 1: Acoustic structure design method based on the deep learning.** (a) The proposed deep learning approach can realize the on-demand design of the acoustic structures. In this work, the deep learning approach is used to design the THRs. The THRs are placed as a side branch of a tube, which can be used to filter out the background noise and make the speech clearer. (b) Schematic view of the THR: The THR is constructed by neck-and-cavity substructures of two elements arranged in a cascade way. When the THR is placed as a side branch of a tube, there are two STL peaks corresponding to the two resonant frequencies $f_1$ and $f_2$. (c) Schematic view of the DNN model. The THRs are designed by a trained DNN. The DNN has a cascaded structure of three hidden layers of nonlinear processing units, where each layer uses the output from the previous layer as input.

Owing to the excellent characteristics of manipulating low-frequency sound waves with subwavelength dimensions, acoustic structures utilizing Helmholtz resonators have emerged as an attractive option in various fields such as sound proofing [35,36], asymmetric sound transmission [37], sound metadiffusers [38], and acoustic superlens [39,40] et al. Moreover, Helmholtz resonators offer significant flexibility and scalability, often being modified and combined for practical applications. For example, Liu et al. proposed a metasurface muffling coating for pipeline ventilation noise reduction, utilizing a Fabry-Perot channel composite design to form a Helmholtz resonance cavity, achieving low-frequency broadband noise reduction [41]. However, conventional Helmholtz resonators are limited to supporting a single monopolar resonant mode with a narrow bandwidth, which restricts their use in functional device design. To address this limitation, recent studies have introduced multi-order Helmholtz resonators capable of producing multiple resonances [42–45]. Compared to conventional Helmholtz resonators, multi-order Helmholtz resonators can provide additional resonances at higher frequencies without significantly increasing volume. For example, Liu et al. achieve multi-order sound absorption by inserting one or more separating plates with small holes into the interior of a Helmholtz resonator, effectively broadening the absorption bandwidth significantly [46]. Figure 1b illustrates a schematic of a two-order Helmholtz resonator (THR), which is assembled from two neck-and-cavity substructures arranged sequentially. The lower substructure is the 1$^{st}$-order element, and the upper substructure is the 2$^{nd}$-order element. Here, $a_i$ and $r_i$ are the radii of the cylindrical necks and cavities, respectively; $l_i$ and $h_i$ are the lengths of the cylindrical necks and cavities, respectively.

The acoustic properties of the THR can be analyzed by the LPT exactly. The $i^{th}$-order cavity is equivalent to the acoustic compliance $C_i$. The $i^{th}$-order neck is equivalent to the acoustic inertance $M_i$ and the acoustic resistance $R_i$.The relationship between GPs $\mathbf{gp} = [a_1, l_1, r_1, h_1, a_2, l_2, r_2, h_2]$ and EEPs $\mathbf{eep} = [R_1, M_1, C_1, R_2, M_2, C_2]$ is shown in Table 1. Here, $V_i = \pi r_i^2 h_i$ is the volume of the $i^{\text{th}}$-order cavity; $\rho_0$ is the static air density; $c_0$ is the sound speed, and $\eta$ is the viscosity of air. $\delta_i = \frac{8a_i}{3\pi}(2 - \beta_i \frac{a_i}{r_i})$ is the end correction associated with the inner and outer openings of the $i^{\text{th}}$-order neck, where $\beta_1 = 0.75$ and $\beta_2 = 1.05$ are the correction factors ($i = 1,2$). The values of the physical constants can be seen in Table S1 of the Supplementary Materials. Based on these EEPs, the acoustic impedance of the THR can be written as follows:

$$Z_{THR} = R_1\sqrt{\omega} + \mathrm{j}\omega M_1 + \cfrac{1}{\mathrm{j}\omega C_1 + \cfrac{1}{R_2\sqrt{\omega} + \mathrm{j}\omega M_2 + 1/\mathrm{j}\omega C_2}}, \tag{1}$$

where $\omega = 2\pi f$ is the angular frequency. When the THR is placed as a side branch of a tube as shown in Figure 1a, the sound transmission loss (STL) can be expressed as:

$$t = 10\log\frac{X_b^2 + \left(\frac{Z_0}{2S} + R_b\right)^2}{R_b^2 + X_b^2}, \tag{2}$$

where $R_b$ and $X_b$ are the real part and imaginary part of $Z_{THR}$, respectively; $Z_0 = \rho_0 c_0$ is the acoustic impedance of the air; $S$ is the cross section of the tube [47]. Considering that the THR with two neck-and-cavity substructures can induce two discrete resonant modes, there are two peaks in the STL curve as shown in Figure 1b. The frequency corresponding to the low-frequency peak is 1st-order resonant frequency $f_1$, and the frequency corresponding to the high-frequency peak is 2nd-order resonant frequency $f_2$. Therefore, the forward problem of the THR model can be solved satisfactorily through the LPT.

| $GP \rightarrow EEP$ | $EEP \rightarrow GP$ |
|---|---|
| $R_i = l_i\sqrt{2\eta\rho_0}/\pi a_i^3$ | $\frac{\rho_0 R_i}{\sqrt{2\eta\rho_0}} a_i^2 - \left(\frac{8\rho_0\beta_i}{3r_i\pi^2} + M_i\right) a_i + \frac{16\rho_0}{3\pi^2} = 0$ |
| $M_i = \rho_0(l_i + \delta_i)/\pi a_i^2$ | $l_i = \frac{3\pi R_i a_i}{\sqrt{2\eta\rho_0}}$ |
| $C_i = V_i/\rho_0 c_0^2$ | $V_i = C_i\rho_0 c_0^2$ |

**Table 1. Relationship between GPs and EEPs**.

### 2.2 Challenge of the inverse design problem

In practical applications such as sound insulation, acoustic structures are usually designed according to noise spectra. And we want to find a structure whose STL values are high enough at the noise frequencies. Equation (2) shows that STL reaches its maximum value when the imaginary part $X_b = 0$, and the corresponding frequencies are the resonant frequencies of THR. Therefore, the resonant frequencies are generally designed closer to the noise frequencies to have good performance of sound isolation. If we let $X_b = 0$ and simplify the equation, a six-degree equation can be obtained:

$$\begin{aligned} M_1 C_1^2 M_2^2 \omega_0^6 + M_1 C_1^2 R_2^2 \omega_0^5 - C_1 M_2 \left(2M_1 \tfrac{C_1+C_2}{C_2} + M_2\right) \omega_0^4 - C_1 R_2^2 \omega_0^3 \\ + \left[M_1 \left(\tfrac{C_1}{C_2}\right)^2 + 2(M_1 + M_2)\tfrac{C_1}{C_2} + M_1 + M_2\right] \omega_0^2 - \tfrac{C_1+C_2}{C_2^2} = 0, \end{aligned} \tag{3}$$

where $\omega_0$ is the resonant angular frequency. According to Table 1 and Eq. (3), it can be seen that all the six GPs $[a_1, l_1, h_1, a_2, l_2, h_2]$ are strong-coupling and affect the resonant characteristics of the THR together, which bring difficulties to accurate inverse design. When we adjust multiple GPs at the same time, the effect on the acoustic functionality of THR is irregular. Therefore, it is difficult to design THR for a target acoustic functionality through manual design method.

Considering of the difficulty of inverse design, DNN is suitable to assist in the design process for avoiding time-consuming numerical methods. However, establishing a relationship between target acoustic functionalities and acoustic structures is often a one-to-many multivalued problem, posing a significant challenge for DNN construction and training. First of all, there can be multiple STL spectra that meet target acoustic functionalities, each corresponding to a different set of EEPs. This implies that multiple sets of EEPs can meet the design requirements. Secondly, a single set of satisfactory EEPs also corresponds to more than one acoustic structures. Figure 2 shows this multivalued challenge. Here, the design target is to find a structure whose resonant frequencies are 150 Hz and 250 Hz, where the values of the STL are at least 10 dB. Four satisfied structures are shown in Figure 2. All of the STL spectra of these four structures meet the requirements, which are shown in Figure 2a and c. Structure 1 and Structure 2 correspond to the same group of EEPs, whose schematic views and GPs are shown in Figure 2b. Structure 3 and Structure 4 correspond to another group of EEPs, whose schematic views and GPs are shown in Figure 2d. It can be seen that the actual THR structure can vary quite significantly for the same design requirement. If the target acoustic functionalities and GPs of the acoustic structures are treated as inputs and outputs for the DNN model, respectively, the training dataset may contain numerous conflicting samples with identical inputs but varied outputs. This scenario, where the DNN is intended to create a one-to-one mapping between inputs and outputs, can complicate the convergence of the training.

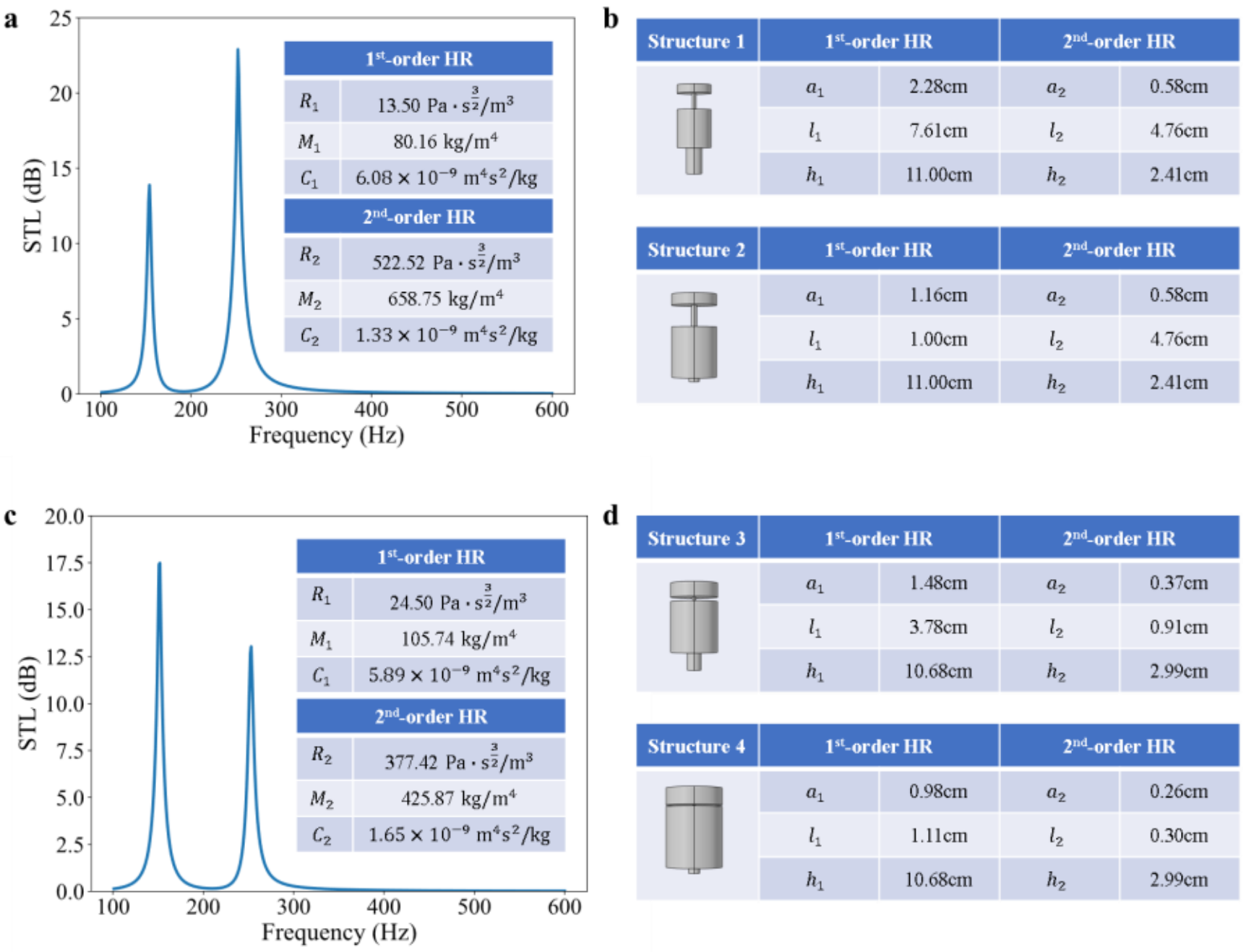

**Figure 2: Different acoustic structures corresponding to the same target acoustic functionality.** (a) STL spectrum calculated by LPT, whose corresponding EEPs are shown in the inset. (b) The two THR structures corresponding to the EEPs shown in (a). (c) STL spectrum calculated by LPT, whose corresponding EEPs are shown in the inset. (d) The two THR structures corresponding to the EEPs shown in (c).

## 2.3 DNN model

Our goal is to use a DNN model to solve the inverse design problem for the target acoustic functionality. As mentioned in the previous section, the mapping from target acoustic functionalities to acoustic structures is typically one-to-many, which brings difficulties to model training. Therefore, some strategies need to be introduced to minimize the one-to-many effect. Firstly, the DNN model is trained to learn the relationship between STL spectra and EEPs, instead of the relationship between STL spectra and GPs. In this way, the one-to-many multivalued problem, which is caused by the one-to-many mapping between EEPs and GPs, can be avoided. Moreover, we make the STL spectrum, instead of the target resonant frequencies, be the input of the DNN model. Two different groups of EEPs may have the same values of the STL at some frequency points, but it is unlikely to be the same at all of the frequency points. With the help of the above two strategies, the model training can converge successfully. And the EEPs, which would most closely produce the target STL spectrum, can be predicted based on the trained DNN model. The GPs can be calculated based on the EEPs through the LPT (see Table 1).

The datasets were generated using LPT. Here, the ranges of the GPs are set as $0.1\ \text{cm} < a_i < 2.5\ \text{cm}$, $0.1\ \text{cm} < l_i < 5\ \text{cm}$ and $0.1\ \text{cm} < h_i < 12.7\ \text{cm}$ ($i = 1,2$), and the radii of the cavities are set as $r_1 = r_2 = 5cm$. The cross section of the tube is $S = 100\ \text{cm}^2$. To generate a sample, first, we sampled randomly within the given ranges of the GPs and obtained a group of GPs. Next, the EEPs and the STL spectrum could be calculated through the LPT. If the STL at the resonant frequencies exceeded 10 dB, the sample was retained. The above selection guarantees that all samples in the dataset have strong sound isolation performance at the resonant frequencies, which contributes to developing a more effective model. Furthermore, we need to ensure that the samples were uniformly distributed across the frequency range of interest. To achieve this, the samples were divided into 35 groups based on their resonant frequencies, ensuring that the distribution of the samples is as uniform as possible (see Table S2 in the Supplementary Materials). Figure 3a shows the distribution of the filtered samples. Here, $f_1$ and $f_2$ are the 1st-order and 2nd-order resonant frequencies of the THR, respectively. The two resonant frequencies $f_1$ and $f_2$ of the samples in the datasets cover the frequency range of interest. Each square in this figure is corresponding to a group of samples in the certain frequency band, where the center frequencies are $f_1$ and $f_2$, and the bandwidth is 50 Hz. The color of the square represents the average STL at the resonant frequency $\frac{t(f_1)+t(f_2)}{2}$ of the samples.

In total, our dataset contains 195,000 samples, which were split into training, validation, and test sets, with proportions of 80%, 10%, and 10%, respectively. We trained a fully connected network with three hidden layers based on this dataset. The network's structure is illustrated in Figure 1c. A desired STL spectrum $\mathbf{t} = [t_1, t_2, t_3, \ldots, t_n]$ is taken as the input, where the STL values are higher than 10 dB at the target frequencies. The STL spectrum is sampled from 101 Hz to 600 Hz with a step of 1 Hz. Thus, the number of inputs is $n = 500$. To facilitate model training, the EEPs of the samples were normalized to the following ranges: $1\ \mathrm{Pa \cdot s^{\frac{3}{2}}/m^3} < R_i < 170\ \mathrm{Pa \cdot s^{\frac{3}{2}}/m^3}$ , $1\ \mathrm{kg/m^4} < M_i < 300\ \mathrm{kg/m^4}$ and $7 \times 10^{-10}\ \mathrm{m^4s^2/kg} < C_i < 7 \times 10^{-9}\ \mathrm{m^4s^2/kg}$ $(i = 1,2)$. The model's output is the normalized 6-dimensional EEPs. The numbers of neurons of each hidden layer are 400, 250 and 220 respectively.

The inputs were normalized, shuffled and then fed into the network, which can accelerate convergence of the algorithm. The mean square error (MSE) was used to represent a loss function between the desired and actual output. The train loss was used to generate the gradients, and the network weights were updated by the Adam algorithm to minimize the discrepancy [48]. The hyperparameters (for example, number of hidden layers, neurons and learning rate) were set according to the performance on the validation set through grid search. We utilized the batch normalization technique to improve the convergence speed of the training [49]. In addition, the dropout regularization technique was employed to avoid overfitting [50]. The hyperparameters for model training can be found in Table 2. The learning rate is set as 0.001, and the batch size is set as 256. We stopped training when the validation loss stopped decreasing, and the learning curve of the DNN on the validation set is shown in Figure 3b. The rapidly decreasing MSE of validation instances shows that training is highly effective.

The test set is used to evaluate the generalization capability of the trained DNN. The average loss of the test set converges to 0.0029. And the prediction results on the test set can be found in S1 of the Supplementary Materials. Based on these results, it can be concluded that the predictive error is within an acceptable range, indicating that the model possesses satisfactory generalization performance. Considering that our model contains 358,948 trainable parameters and the STL spectrum is highly sensitive to variations in EEPs, a large amount of high-quality data is essential to ensure the model's performance and prevent overfitting. Detailed analysis of the high-quality data selection and the dataset size can be found in S2 and S3 of the Supplementary Materials, respectively.

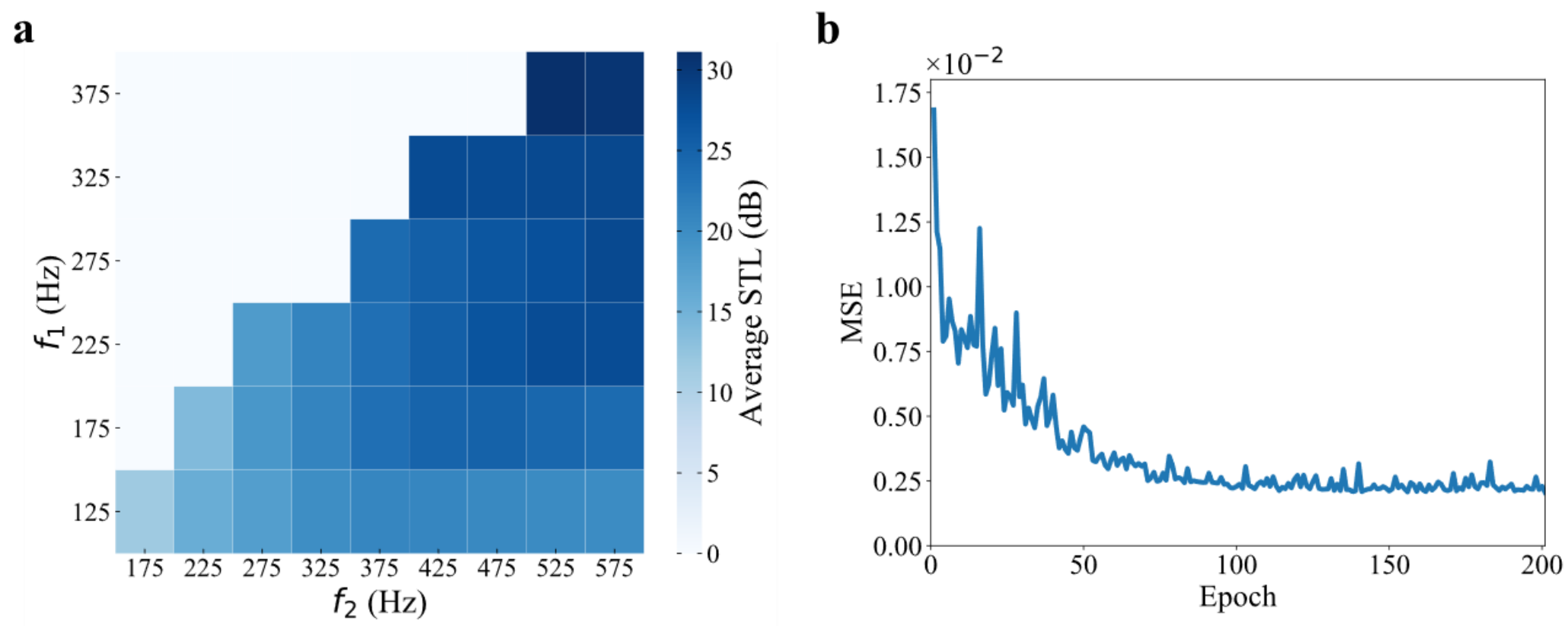

**Figure 3: Distribution of samples and learning curve of the DNN.** (a) Distribution of the filtered samples. (b) The MSE of validation set over the epoch.

| Learning Rate | Batch Size | Dropout Rate | Activation function |
|---|---|---|---|
| 0.001 | 256 | 0.2 | ReLu |

**Table 2. Hyperparameters for model training.**

## 3. Results

### 3.1 DNN solves the inverse design problem

The concept schematic of the design process is shown in Figure 4. Here, we show an example using the trained DNN to realize acoustic insulation at specific frequencies. The target resonant frequencies are set as $f_1^{\text{target}} = 150\text{Hz}$ and $f_2^{\text{target}} = 250\text{Hz}$, where the values of the STL are required to exceed 10 dB. Then we generate a group of curves as the desired STL spectra according to the requirements. Considering that the profile of the STL peak is similar to the Lorentzian profile, a desired STL curve can be created by scaling, translating and overlaying the standard Lorenz curve. Although it cannot be guaranteed that an STL spectrum generated in this manner will definitely correspond to a group of EEPs, the proposed DNN model, with its robust fitting capabilities, will endeavor to predict the EEPs that best meet the requirements. We feed these spectra into the proposed DNN model and obtain the predictive EEPs. Then, the GPs can be calculated through the LPT. Finally, we can calculate the real STL spectra by the TMM (see S4 of the Supplementary Materials) and select the most qualified structure based on their average error of the resonant frequency (AERF), which can be expressed as

$$\mathrm{AERF}(\mathbf{gp}) = \frac{1}{2}\left(\frac{\left|f_1(\mathbf{gp}) - f_1^{\mathrm{target}}\right|}{f_1^{\mathrm{target}}} + \frac{\left|f_2(\mathbf{gp}) - f_2^{\mathrm{target}}\right|}{f_2^{\mathrm{target}}}\right). \tag{4}$$

Here, $f_1(\mathbf{gp})$ and $f_2(\mathbf{gp})$ are the real resonant frequencies of the structure whose GPs are denoted as $\mathbf{gp}$. By using TMM to refine the predicted structures, we can reduce the approximation error introduced by the LPT, ensuring that the designed THR better meet the design requirements. Moreover, if computational resources allow, more accurate methods such as finite element analysis (FEM) can be used for further refinement and selection of the best design.

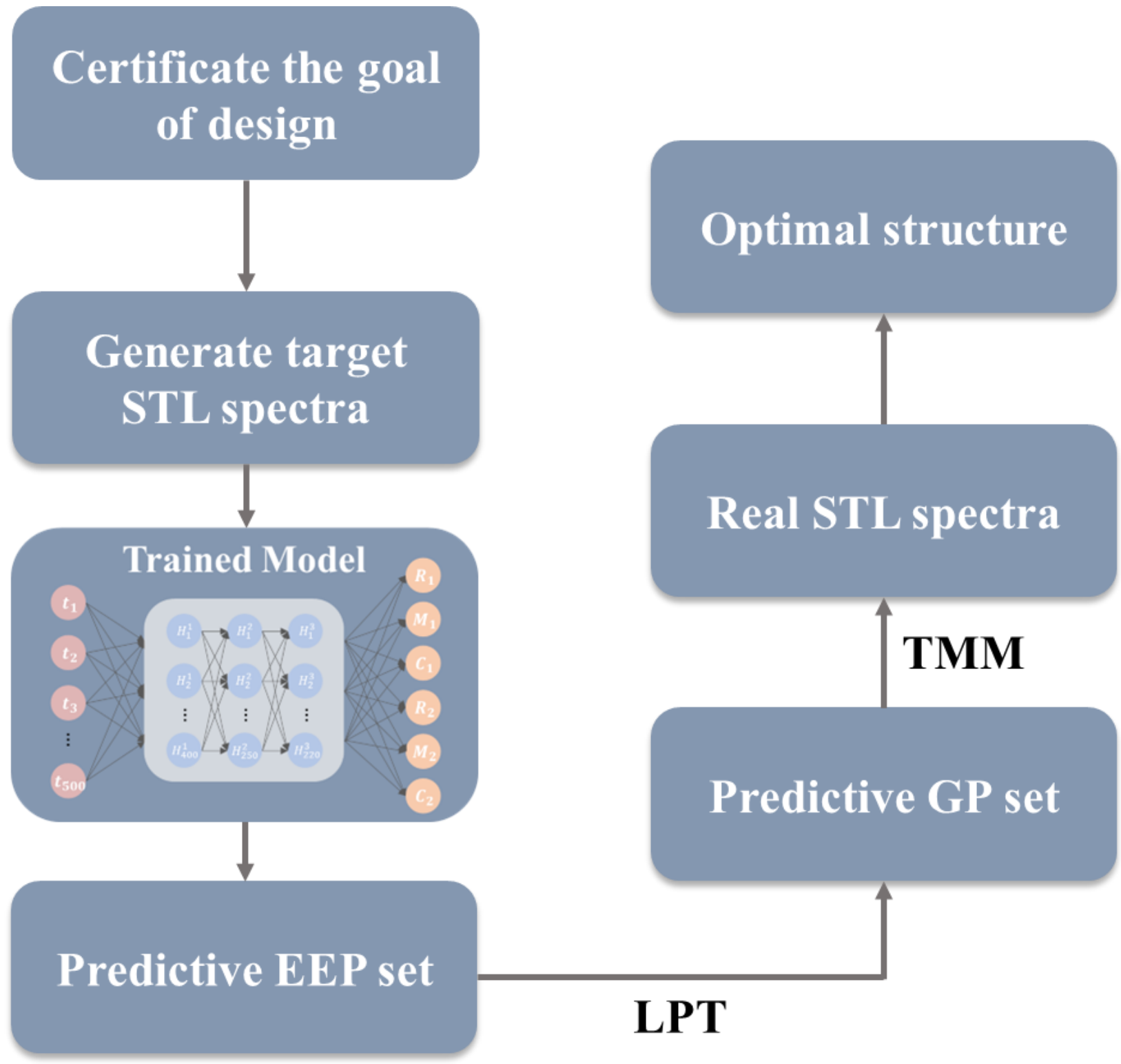

**Figure 4: Concept schematic of the design process.**

Based on the aforementioned methodology, the selected structure $\beta_0$ is produced, whose GPs and STL spectrum are shown in Figure 5a. There are two STL peaks higher than 10 dB at 150 Hz and 250 Hz, which is in good agreement with the design requirements. Based on the proposed model, we can further implement quantitative design to specify STL values, such as 9dB or 12dB, at target frequencies, which is highly valuable for scenarios that require more precise designs, such as the development of acoustic filters [6, 51]. The results can be found in S5 of the Supplementary Materials.

Building on this foundation, we introduced minor perturbations to the GPs of $\beta_0$, resulting in significant changes to the STL spectrum, which no longer met the design requirements (see S6 of the Supplementary Materials). This further demonstrates that the inverse design problem necessitates a refined adjustment of the GPs. The proposed design scheme based on DNN has proven to be highly effective in addressing this issue. In addition to the aforementioned, we have also chosen design objectives that were not encompassed within the training dataset to assess the generalization capability of the DNN model. The outcomes can be found in S7 of the Supplementary Materials.

It is noteworthy that the model directly predicts the EEPs of the acoustic structure without imposing any constraints on the geometric structure itself. The THR is merely one physical realization of this set of EEPs. Due to the limitations of the mapping relationship between GPs and EEPs, there may be instances where the EEPs predicted by the model cannot be realized by a specific THR. In such cases, we can seek alternative acoustic structures capable of realizing this set of EEPs within the framework of LPT. In S8 of the Supplementary Materials, we further demonstrate the flexibility and transferability of the proposed model using the double-neck THR as an illustrative example.

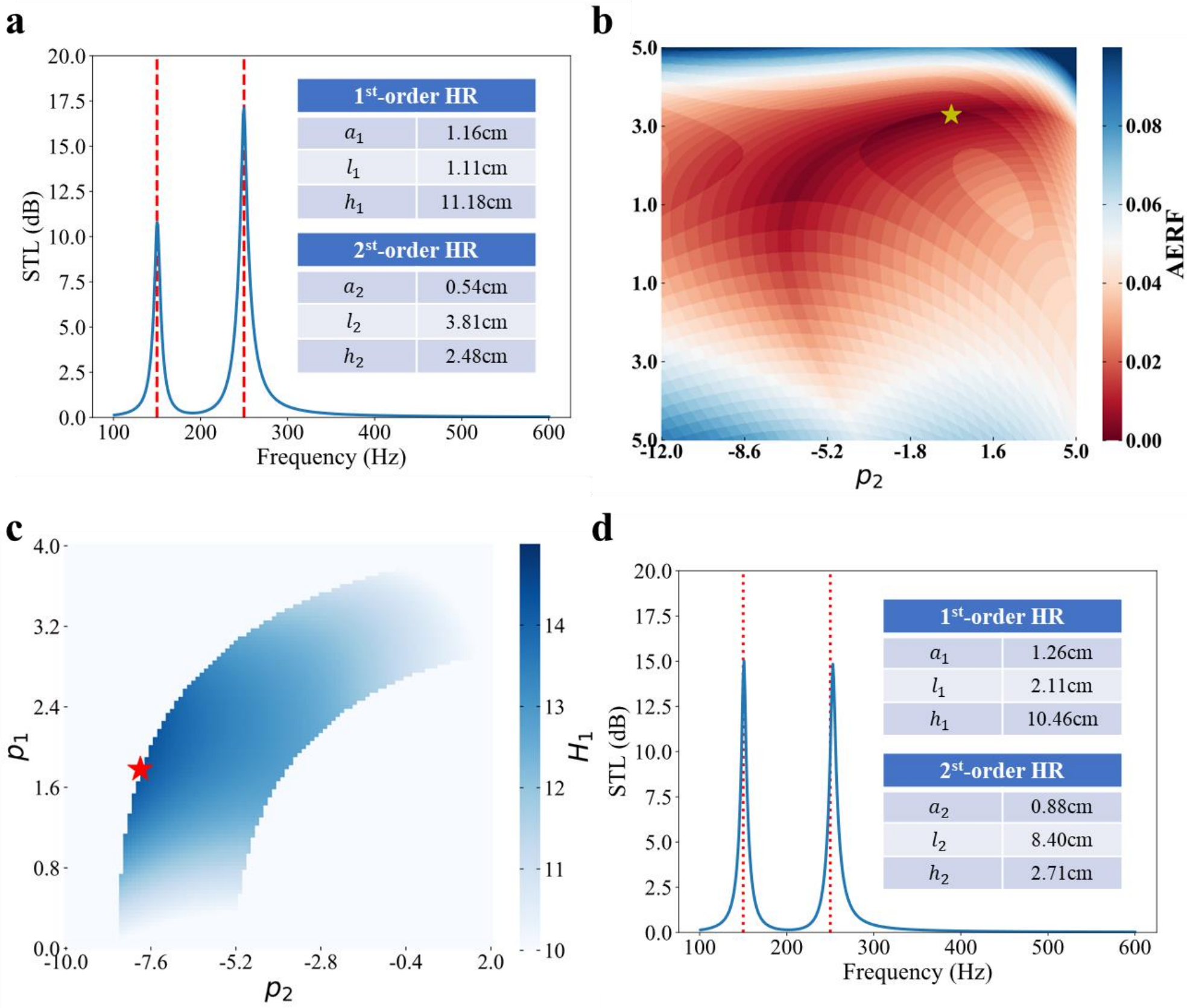

**Figure 5: Results of the inverse design**. (a) STL spectrum of the selected THR structure $\beta_0$ calculated by the TMM (blue solid line). The target resonant frequencies 150 Hz and 250 Hz are marked by the red dotted lines. The GPs of the selected THR structure are shown in the table. (b) AERF varies with the two most important principal components $p_1$ and $p_2$ in PCA space. The selected structure $\beta_0$ is marked with the yellow star. (c) $H_1$ varies with the two most important principal components $p_1$ and $p_2$. The best structure $\beta_1$ is marked with the red star, whose STL at 150 Hz 14.26 dB. (d) STL spectrum of the structure $\beta_1$, whose GPs are shown in the table.

## 3.2 Search for multiple solutions

As mentioned before, the corresponding relation between target acoustic functionality and acoustic structures is usually not a deterministic one-to-one mapping. In the previous section, the design target is to find a structure whose resonant frequencies are 150 Hz and 250 Hz, where the values of the STL are at least 10 dB. This is a one-to-many multivalued problem. Considering that the multiple solutions distribute in the whole GP space, it will take a prohibitive amount of time to find them through searching in GP space. Here, the proposed DNN model and the principal component analysis (PCA) can be combined to search for multiple solutions quickly. PCA is a linear transformation algorithm, which can find a sequence of best orthogonal linear projections (called principal components) that maximizes the corresponding variances. Therefore, PCA can be used to map and characterize the multi-parameter design space, which can transform a set of correlated variables into a set of new uncorrelated variables [52].

The PCA variable space can be defined by six orthogonal basis vectors $\mathbf{V_1}$, $\mathbf{V_2}$, …, $\mathbf{V_6}$. In the PCA variable space, the selected structure $\beta_0$ shown in Figure 5a can be described as $\mathbf{P}^{\beta_0} = \left[p_1^{\beta_0}, p_2^{\beta_0}, \dots, p_6^{\beta_0}\right]$, and the structures with similar acoustic properties will gather together. Therefore, we can search around the $\mathbf{P}^{\beta_0}$ to find other satisfied solutions. In order to visualize the results, we generate a 2D hyperplane defined by the two most important principal components $\mathbf{V_1}$ and $\mathbf{V}_2$. For each point $\left[p_1^i, p_2^i\right]$ on the hyperplane, the corresponding GPs $\mathbf{gp}^i = \left[a_1^i, l_1^i, r_1^i, h_1^i, a_2^i, l_2^i, r_2^i, h_2^i\right]$ in the original design space can be computed as $\mathbf{gp}^i = \mathbf{V}_0 + p_1^i\mathbf{V}_1 + p_2^i\mathbf{V}_2 + p_3^{\beta_0}\mathbf{V}_3 + p_4^{\beta_0}\mathbf{V}_4 + p_5^{\beta_0}\mathbf{V}_5 + p_6^{\beta_0}\mathbf{V}_6$, where $p_1^i$ varies from -5 to 5 and $p_2^i$ varies from -12 to 5.

And the corresponding AERFs are computed and shown in Figure 5b. It is clear that the selected structure $\beta_0$ (marked with a yellow star) is one of the optimal solutions within the range of observations, which demonstrates the accuracy of the proposed design technology.

In fact, there are many points surround $\beta_0$ in PCA space whose corresponding acoustic properties are also in good agreement with the design requirements. However, the corresponding structures of these points may vary considerably. Therefore, we can make further selections based on different properties in other aspects. For example, if the target is to obtain a structure whose STL in 150 Hz can be as high as possible and STL in 250 Hz can higher than 10 dB, we can define $H_1$ as

$$H_1 = \begin{cases} t(f_1^{\text{target}}, \mathbf{gp}) & if\ t(f_1^{\text{target}}, \mathbf{gp}), t(f_2^{\text{target}}, \mathbf{gp}) > 10dB \\ 0 & else \end{cases}, \quad (5)$$

where $f_1^{\text{target}} = 150\ Hz$, $f_2^{\text{target}} = 250\ Hz$; $t(f_1^{\text{target}}, gp)$ are the values of the STL at 150 Hz of the THR, whose GPs are denoted as $\mathbf{gp} = [a_1, l_1, r_1, h_1, a_2, l_2, r_2, h_2\ ]$. It can be seen that the higher $H_1$ is, the better the corresponding structure meet the design requirements. Then, $H_1$ of the points surround $\beta_0$ can be calculated. The results are shown in Figure 5c. We pick the point with highest $H_1$ (marked with red star) within the range of observations, which corresponding structure is $\beta_1$. The STL spectrum and GPs of the structure $\beta_1$ are shown in Figure 5d. It can be seen that the STL at 150 Hz is improved to 14.26 dB compared with the structure $\beta_0$. Details about the PCA and another example with high performance in 250 Hz are provided in S9 of the Supplementary Materials.

Consequently, a lower-dimensional subspace that exhibits high performance can be discerned utilizing dimensionality reduction methodologies. Within this reduced subspace, the predictive outcomes furnished by the DNN model serve as an anchor point. By conducting a localized search in the vicinity of this anchor, we are enabled not only to conduct a more efficient search for multiple solutions but also to make additional selections with a reduced computational cost when various performance criteria are taken into account.

### 3.3 DNN aided optimization method

In the previous section, our goal was to design a structure with STL values that surpassed 10 dB at the target frequencies. Yet, in real-world scenarios, the higher these values, the better, often requiring the application of optimization algorithms to adjust the structural parameters effectively [53]. A popular approach involves the use of evolutionary algorithms, such as genetic algorithms (GA) and particle swarm optimization [54–56]. However, these algorithms can be quite sensitive to the starting conditions, which may result in challenges like slow convergence or getting stuck at local optima. Now that we can easily design an arbitrary STL spectrum with the proposed DNN, we can leverage it to offer a solid starting point for the evolutionary algorithms. Specifically, the initial population could consist of a set of high-performing individuals generated by the DNN. This step can be seen as introducing prior knowledge to the optimization process, enhancing the algorithms' effectiveness from the outset.

Here, we want to maximize the average value of the STL at the target frequencies, which can be recast as

$$\min_{\mathbf{gp}} J = \min_{\mathbf{gp}} -[t(f_1^{\text{target}}, \mathbf{gp}) + t(f_2^{\text{target}}, \mathbf{gp})]$$
$$s.t.\ t(f_1^{\text{target}}, \mathbf{gp}) > 10 \text{ dB} \text{ and } t(f_2^{\text{target}}, \mathbf{gp}) > 10 \text{ dB}, \quad (6)$$

where $f_1^{\text{target}}$ and $f_2^{\text{target}}$ are still set as 150 Hz and 250 Hz, respectively; $t(f_1^{\text{target}}, \mathbf{gp})$ and $t(f_2^{\text{target}}, \mathbf{gp})$ are the values of the STL at the target frequencies of the THR, whose GPs are denoted as $\mathbf{gp} = [a_1, l_1, r_1, h_1, a_2, l_2, r_2, h_2]$. The ranges of GPs are same with those of the datasets for model training. The objective function $F(f_1^{\text{target}}, f_2^{\text{target}}, \mathbf{gp})$ can be expressed as:

$$F = \begin{cases} -[t(f_1^{\text{target}}, \mathbf{gp}) + t(f_2^{\text{target}}, \mathbf{gp})] & if\ t(f_1^{\text{target}}, \mathbf{gp}), t(f_2^{\text{target}}, \mathbf{gp}) > 10dB \\ \lambda - [t(f_1^{\text{target}}, \boldsymbol{gp}) + t(f_2^{\text{target}}, \mathbf{gp})] & else \end{cases}, \quad (7)$$

where $\lambda$ is the penalty parameter, which is expected to help the real resonant frequencies of the individuals $f_1$ and $f_2$ to get closer to $f_1^{\text{target}}$ and $f_2^{\text{target}}$ during the evolution. Therefore, the penalty parameter $\lambda$ can be set as:

$$\lambda = \begin{cases} 10 + |f_1 - f_1^{\text{target}}| + |f_2 - f_2^{\text{target}}|, & if\ 101 \leq f_1, f_2 \leq 600 \\ 1000, & else \end{cases}. \quad (8)$$

The objective function $F(f_1^{\text{target}}, f_2^{\text{target}}, \mathbf{gp})$ was minimized through the GA, which was realized by an open-source evolutionary algorithm toolbox and framework for Python, Geatpy [57]. The soea_SEGA module of Geatpy with the default algorithmic parameters was applied in this task.

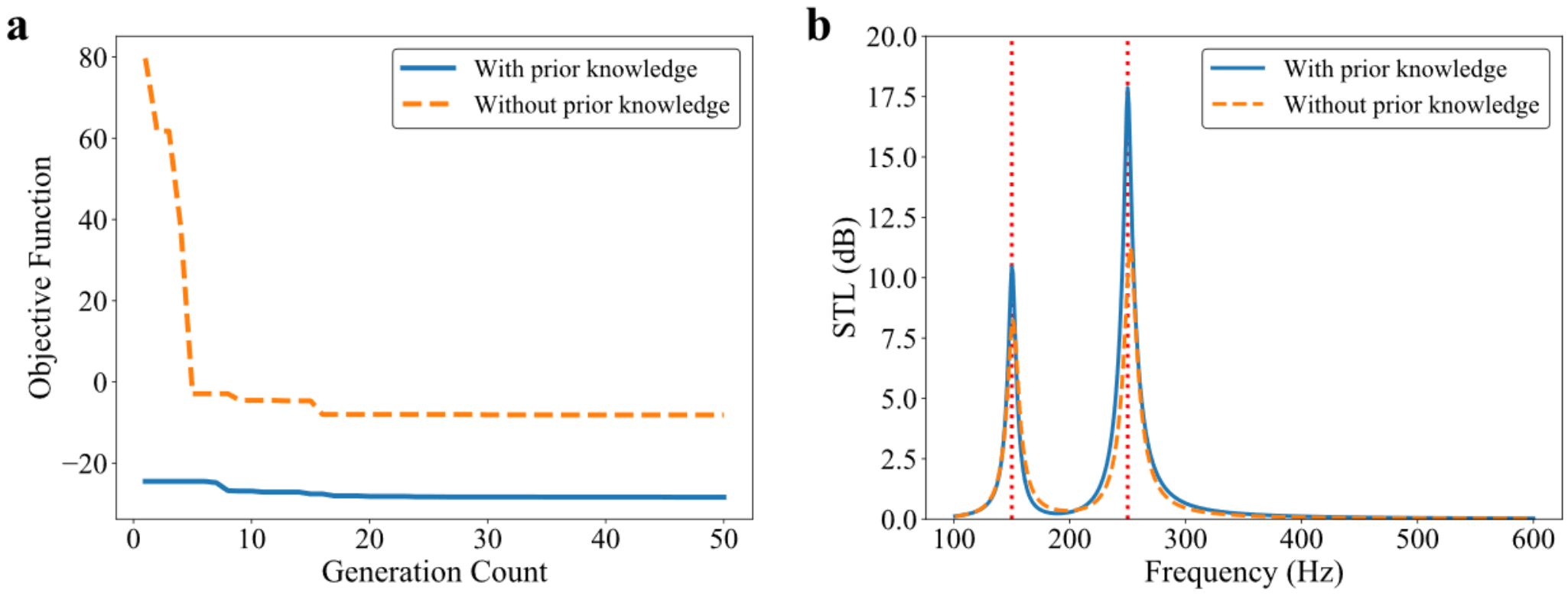

**Figure 6: The optimization effect through the GA with and without prior knowledge.** (a) Comparison of the decreasing process of the objective function $F$ under the two conditions (with prior knowledge, blue solid line; without prior knowledge, yellow dotted line). (b) Comparison of the STL spectra of the optimal structures under the two conditions (with prior knowledge, blue solid line; without prior knowledge, yellow dotted line).

The optimizations were conducted using GA under two different initial conditions. In the first scenario, the initial population comprised 50 individuals, all randomly generated, indicating that we started with no prior knowledge about this task. In the second scenario, the initial population was a mix of 45 randomly generated individuals and 5 high-performing elitist individuals. These elitist individuals, with resonant frequencies of 150 Hz and 250 Hz, were generated by our proposed DNN model. Each elitist individual can be obtained through a simple query based on the DNN that takes just a few milliseconds. To ensure a fair comparison, all other algorithmic parameters were kept consistent. The algorithm reached convergence after 50 generations. The comparative effectiveness of the optimizations under these two conditions is illustrated in Figure 6. Figure 6a shows the comparison of the decreasing process of the objective function $F$. Considering that the initial population with prior knowledge includes a certain number of elitist individuals, the blue solid line converges faster than the yellow dotted line. Figure 6b shows the comparison of the STL spectra of the optimal structures under the two conditions. It can be seen that the average STL at the target resonant frequencies with prior knowledge (14.2 dB) is higher than the average STL at the target resonant frequencies without prior knowledge (9.6 dB). The above results demonstrate that a good initial condition can not only effectively accelerate convergence velocity, but also improve the results of the optimization. Therefore, the DNN-aided optimization method is feasible and more efficient for optimizing the acoustic structure.

### 3.4 On-demand design of the acoustic filter

In practical applications, we need to filter out the multi-frequency line-spectrum noise in the background environment, and even achieve broadband sound insulation in a certain frequency range, requiring us to design a combined filter for specific noise frequencies. The traditional design method needs to perform the optimization for each noise frequency, so the efficiency is very low. In comparison, it takes only a few seconds to complete the design process using our design strategy. Here, we demonstrate an example of designing an acoustic filter to decrease the line-spectrum noise at four frequencies: 150 Hz, 200 Hz, 250 Hz and 300 Hz. To realize four resonant frequencies, the acoustic filter is a combination of two THRs. The two THRs were designed using the DNN model, with the photo and GPs shown in Figure 7a. The STL spectra of the acoustic filter are shown in Figure 7b, where the theoretical (blue dotted line, calculated by the TMM), simulated (orange solid line, calculated by the FEM) and experimental (green circles) results are consistent (see Appendix A for details). There are four transmission loss peaks at 150 Hz, 200 Hz, 250 Hz and 300 Hz, corresponding to the four resonant modes of the two THRs. These results confirm the design is very precise.

To evaluate the effectiveness of the acoustic filter, a pure voice mixed with noise at the above frequencies impinges from the left port of a square tube. A microphone is used to receive the signal at the right port of the waveguide. We performed the measurements with and without acoustic filter as a side branch of the tube. The time-domain waveform and the spectrogram of the signal filtered by the acoustic filter are shown in Figure 7d and f, respectively. For comparison, the time-domain waveform and the spectrogram of the unfiltered signal are shown in Figure 7c and e, respectively. It can be seen that the proposed acoustic filter significantly decreases the noise energy and improves the speech clarity of the original signal (the sound before and after filtering can be heard in the Supporting Video). And the experimental results prove that our approach can achieve the on-demand design of acoustic filter for many applications, such as noise reduction of the engine, helicopter and UHV transformers.

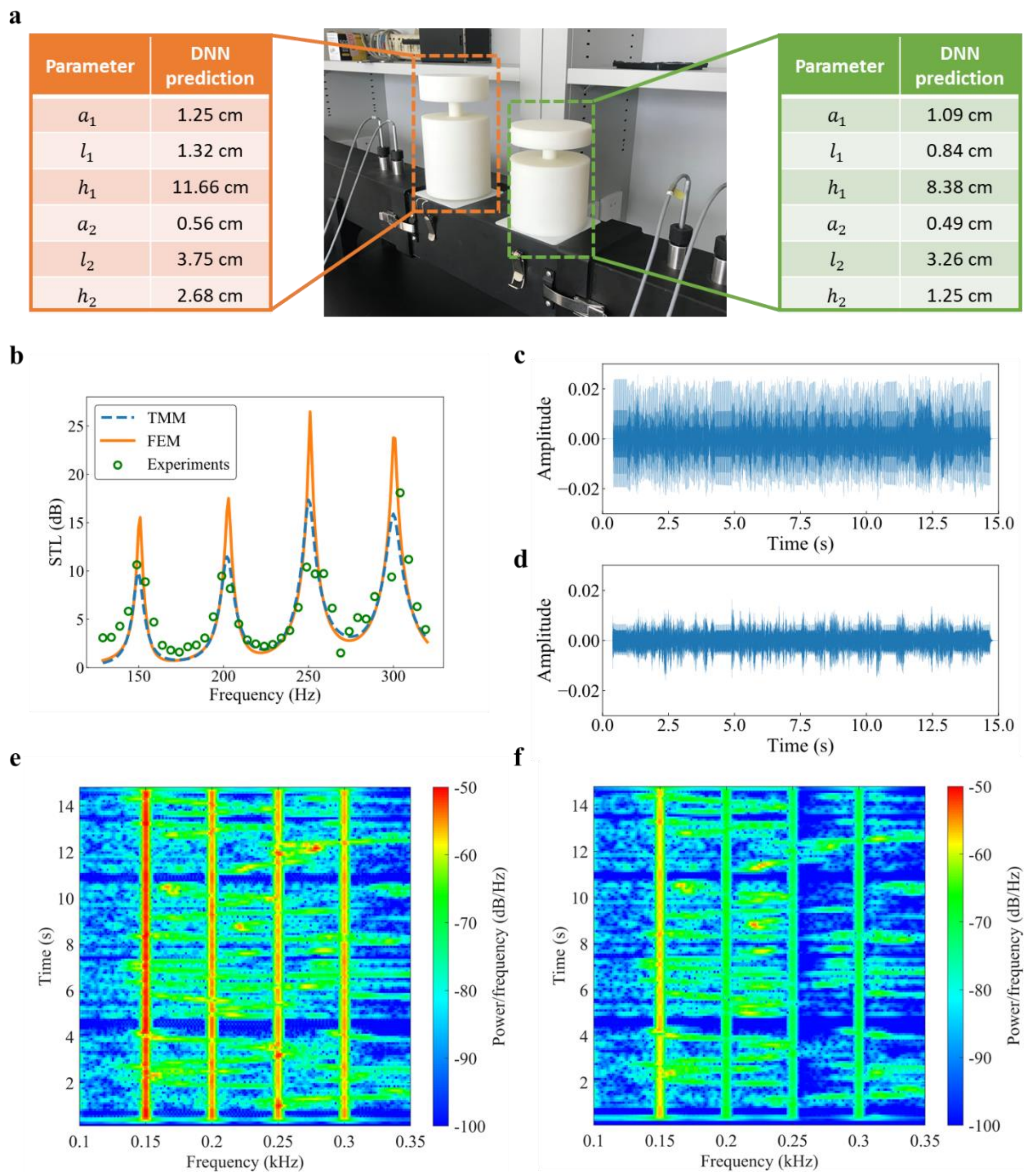

**Figure 7: Structure and experimental results of the acoustic filter**. (a) Photo and GPs of the acoustic filter. Left inset: The GPs of the THR whose resonant frequencies are 150 Hz and 250 Hz. Right inset: The GPs of the THR whose resonant frequencies are 200 Hz and 300 Hz. (b) STL spectrum of the acoustic filter. (c) and (d) are the time-domain waveforms of the test signal before and after filtering, respectively. (e) and (f) are spectrograms of the test signal before and after filtering, respectively.

## 4. Discussion

In this paper, we demonstrate a DNN model for acoustic structure design based solely on desired acoustic properties. The trained model can determine the corresponding relation between the EEPs and acoustic properties, which shows a very accurate prediction of the geometry of acoustic structures with multiple strong-coupling parameters. For the multivalued inverse design problem, we can find one of the satisfied solutions very quickly using the proposed model. Combined with the dimension reduction algorithm, the proposed model also can be used to search for multiple solutions. Moreover, the proposed model is able to aid evolutionary algorithms in completing the optimization task more efficiently. The effectiveness of the structures designed by our approach was evaluated through acoustic experiments.

As mentioned before, the one-to-many relationship between target acoustic functionality and acoustic structures presents a big challenge to DNN building and training. To overcome this, LPT is introduced to the model building, and the proposed model is trained to learn EEPs instead of GPs of the acoustic structures. The combination with LPT is an effective way to avoid using complex network, which needs more training samples and is more difficult to be trained well, to overcome the one-to-many effect [22,58]. Moreover, the proposed model is not limited to solve the inverse design problem of THR. It is able to be easily extended to other acoustic structures, whose equivalent circuit is similar to the THR. Notwithstanding this limitation, the proposed method is certainly more flexible compared with learning the GPs directly as the previous studies.

In conclusion, the proposed deep learning approach shows a significant improvement in efficiency, an acceleration of the design process and an obvious reduction in both computational and man-powered resources. The trained model would be an effective design tool for designers, especially for layman users who have less professional knowledge about acoustics. It should be noted that although this work focuses on inverse design of acoustic structures, the proposed method, which solves the one-to-many multivalued problem through learning equivalent parameters, can also be extended to other research fields where the equivalent circuit approach is frequently used, such as the design of the mechanical, piezoelectric and microwave devices [59–61]. These good generalization performances of the proposed model promise a wide range of potential applications.

## Conflict of Interest

The authors declare no competing interests.

## Authors' Contributions

J.Y. planned, coordinated, and supervised the project. X.S., Y.Y. and H.J. conceived the model. X.S. designed the algorithm and performed the theoretical and numerical calculations. X.S. and H.Z. participated in the experiments and data analysis. X.S., Y.Y. and H.J. jointly wrote the manuscript. All authors contributed to discussing the results and reading the manuscript.

## Acknowledgements

This work is supported by the Key-Area Research and Development Program of Guangdong Province (Grant No. 2020B010190002), the National Natural Science Foundation of China (Grant Nos. 11874383 and 12104480), the IACAS Frontier Exploration Project (Grant No. QYTS202110).

## Appendix A. Methods of numerical simulations and experimental measurements

*A.1. Numerical simulations.*

In this paper, FEM simulations were performed to verify the feasibility of our design by using the pressure acoustic module and thermoviscous acoustic module of COMSOL Multiphysics. Plane wave radiation boundary condition is set on the left side and right side of the calculated fluid domain of the tube. The mesh type is the tetrahedral mesh, and the largest mesh element size was smaller than 1/6 of the shortest incident wavelength, and the further refined meshes were applied in the cylindrical necks. The DNN model is built by PyTorch in Python 3.6. The integrated development environment is Anaconda 3 in Linux operation system. The GPU configuration is NVIDIA GeForce GTX 1080Ti.

*A.2. Experimental measurements.*

The experimental THR samples were fabricated using 3D printing technology with a wall thickness of 5 mm. The material used for the samples is Lasty-KS, a type of UV-curable resin, with density of 1.13 g/cm3. The acoustic impedance of the Lasty-KS is much larger than the acoustic impedance of air, so the wall of the THR can be regarded as rigid for the sound wave.

The layout of the measurement system is shown in Figure 7a. The two THRs were placed as a side branch of a square tube, where the connection was sealed by plasticene. The STL spectrum was measured using four 1/4-inch microphones (Brüel & Kjær, type-4187). The experimental measurements were conducted by the two-load method in a square standing wave tube, where the hard wall and acoustic absorbing sponge (as an anechoic boundary) were separately used at the terminal. When the mixed signals shown in Figure 7c and d were measured, a pure voice mixed with noise impinged from the left port of the square tube. A 1/2-inch microphone (Brüel & Kjær, type-4189) was placed at the right port of the tube to receive the signal, while an acoustic absorbing sponge was used as the anechoic boundary at the right port.

**References**

[1] Y. Yang, H. Jia, Y. Bi, H. Zhao, J. Yang, Experimental Demonstration of an Acoustic Asymmetric Diffraction Grating Based on Passive Parity-Time-Symmetric Medium, Phys Rev Appl. 12 (2019) 034040. https://doi.org/10.1103/PhysRevApplied.12.034040.

[2] Y. Yang, H. Jia, S. Wang, P. Zhang, J. Yang, Diffraction control in a non-Hermitian acoustic grating, Appl. Phys. Lett. 116 (2020) 213501. https://doi.org/10.1063/5.0004104

[3] Y. Bi, H. Jia, Z. Sun, Y. Yang, H. Zhao, J. Yang, Experimental demonstration of three-dimensional broadband underwater acoustic carpet cloak, Appl. Phys. Lett. 112 (2018) 223502. https://doi.org/10.1063/1.5026199

[4] Z. Sun, X. Sun, H. Jia, Y. Bi, J. Yang, Quasi-isotropic underwater acoustic carpet cloak based on latticed pentamode metafluid, Appl. Phys. Lett. 114 (2019) 094101. https://doi.org/10.1063/1.5085568.

[5] N. Wang, C. Zhou, S. Qiu, S. Huang, B. Jia, S. Liu, J. Cao, Z. Zhou, H. Ding, J. Zhu, Y. Li, Meta-silencer with designable timbre, Int. J. Extreme Manuf. 5 (2023) 025501. https://doi.org/10.1088/2631-7990/acbd6d.

[6] X. Sun, H. Jia, Z. Zhang, Y. Yang, Z. Sun, J. Yang, Sound Localization and Separation in 3D Space Using a Single Microphone with a Metamaterial Enclosure, Adv. Sci. 7 (2020) 1902271. https://doi.org/10.1002/advs.201902271.

[7] Z. Sun, Y. Shi, X. Sun, H. Jia, Z. Jin, K. Deng, J. Yang, Underwater acoustic multiplexing communication by pentamode metasurface, J. Phys. D: Appl. Phys. 54 (2021) 205303. https://doi.org/10.1088/1361-6463/abe43e.

[8] F. Ma, Y. Xu, J.H. Wu, Modal displacement method for extracting the bending wave bandgap of plate-type acoustic metamaterials, Appl. Phys. Express 12 (2019) 074004. https://doi.org/10.7567/1882-0786/ab27dd.

[9] G. Hinton, L. Deng, D. Yu, G.E. Dahl, A. Mohamed, N. Jaitly, A. Senior, V. Vanhoucke, P. Nguyen, T.N. Sainath, Deep neural networks for acoustic modeling in speech recognition: The shared views of four research groups, IEEE Signal Process. Mag. 29 (2012) 82–97. https://doi.org/10.1109/MSP.2012.2205597

[10] A. Krizhevsky, I. Sutskever, G.E. Hinton, ImageNet Classification with Deep Convolutional Neural Networks, Commun. ACM 60 (2017) 84–90. https://doi.org/10.1145/3065386

[11] K. Cho, B. van Merriënboer, C. Gulcehre, D. Bahdanau, F. Bougares, H. Schwenk, Y. Bengio, Learning Phrase Representations using RNN Encoder–Decoder for Statistical Machine Translation, Proceedings of the 2014 Conference on Empirical Methods in Natural Language Processing (EMNLP), 2014: pp. 1724–1734. https://doi.org/10.3115/v1/D14-1179.

[12] D. Silver, A. Huang, C.J. Maddison, A. Guez, L. Sifre, G. Van Den Driessche, J. Schrittwieser, I. Antonoglou, V. Panneershelvam, M. Lanctot, Mastering the game of Go with deep neural networks and tree search, Nature 529 (2016) 484–489. https://doi.org/10.1038/nature16961

[13] R. Socher, D. Chen, C.D. Manning, A. Ng, Reasoning With Neural Tensor Networks for Knowledge Base Completion, Advances in Neural Information Processing Systems, Curran Associates, Inc., 2013. https://proceedings.neurips.cc/paper_files/paper/2013/file/b337e84de8752b27eda3a12363109e80-Paper.pdf.

[14] Sanchez-Lengeling Benjamin, Aspuru-Guzik Alán, Inverse molecular design using machine learning: Generative models for matter engineering, Science 361 (2018) 360–365. https://doi.org/10.1126/science.aat2663.

[15] G.B. Goh, N.O. Hodas, A. Vishnu, Deep learning for computational chemistry, J. Comput. Chem. 38 (2017) 1291–1307. https://doi.org/10.1002/jcc.24764

[16] T. Zahavy, A. Dikopoltsev, D. Moss, G.I. Haham, O. Cohen, S. Mannor, M. Segev, Deep learning reconstruction of ultrashort pulses, Optica 5 (2018) 666–673.

https://doi.org/10.1364/OPTICA.5.000666

[17] P. Baldi, P. Sadowski, D. Whiteson, Searching for exotic particles in high-energy physics with deep learning, Nat. Commun. 5 (2014) 4308. https://doi.org/10.1038/ncomms5308.

[18] J. Carrasquilla, R.G. Melko, Machine learning phases of matter, Nat. Phys. 13 (2017) 431–434. https://doi.org/10.1038/nphys4035

[19] A. White, P. Khial, F. Salehi, B. Hassibi, A. Hajimiri, A Silicon photonics computational Lensless Active-flat-optics imaging System, Sci. Rep. 10 (2020) 1689. https://doi.org/10.1038/s41598-020-58027-1

[20] J. Weng, Y. Ding, C. Hu, X.-F. Zhu, B. Liang, J. Yang, J. Cheng, Meta-neural-network for real-time and passive deep-learning-based object recognition, Nat. Commun. 11 (2020) 6309. https://doi.org/10.1038/s41467-020-19693-x.

[21] Y. Rivenson, Z. Göröcs, H. Günaydin, Y. Zhang, H. Wang, A. Ozcan, Deep learning microscopy, Optica 4 (2017) 1437–1443. https://doi.org/10.1364/OPTICA.4.001437

[22] I. Malkiel, M. Mrejen, A. Nagler, U. Arieli, L. Wolf, H. Suchowski, Plasmonic nanostructure design and characterization via Deep Learning, Light Sci. Appl. 7 (2018) 60. https://doi.org/10.1038/s41377-018-0060-7.

[23] A. Sheverdin, F. Monticone, C. Valagiannopoulos, Photonic Inverse Design with Neural Networks: The Case of Invisibility in the Visible, Phys Rev Appl. 14 (2020) 024054. https://doi.org/10.1103/PhysRevApplied.14.024054.

[24] J. Peurifoy, Y. Shen, L. Jing, Y. Yang, F. Cano-Renteria, B.G. DeLacy, J.D. Joannopoulos, M. Tegmark, M. Soljačić, Nanophotonic particle simulation and inverse design using artificial neural networks, Sci. Adv. 4 (2018) eaar4206. https://doi.org/10.1126/sciadv.aar4206

[25] T. Qiu, X. Shi, J. Wang, Y. Li, S. Qu, Q. Cheng, T. Cui, S. Sui, Deep Learning: A Rapid and Efficient Route to Automatic Metasurface Design, Adv. Sci. 6 (2019) 1900128. https://doi.org/10.1002/advs.201900128.

[26] W. Ma, F. Cheng, Y. Xu, Q. Wen, Y. Liu, Probabilistic Representation and Inverse Design of Metamaterials Based on a Deep Generative Model with Semi-Supervised Learning Strategy, Adv. Mater. 31 (2019) 1901111. https://doi.org/10.1002/adma.201901111.

[27] W. Ma, Z. Liu, Z.A. Kudyshev, A. Boltasseva, W. Cai, Y. Liu, Deep learning for the design of photonic structures, Nat. Photonics 15 (2021) 77–90. https://doi.org/10.1038/s41566-020-0685-y.

[28] Y.-T. Luo, P.-Q. Li, D.-T. Li, Y.-G. Peng, Z.-G. Geng, S.-H. Xie, Y. Li, A. Alù, J.

Zhu, X.-F. Zhu, Probability-Density-Based Deep Learning Paradigm for the Fuzzy Design of Functional Metastructures, Research 2020 (2020) 1–11. https://doi.org/10.34133/2020/8757403.

[29] M.J. Bianco, P. Gerstoft, J. Traer, E. Ozanich, M.A. Roch, S. Gannot, C.-A. Deledalle, Machine learning in acoustics: Theory and applications, J. Acoust. Soc. Am. 146 (2019) 3590–3628. https://doi.org/10.1121/1.5133944.

[30] W.W. Ahmed, M. Farhat, X. Zhang, Y. Wu, Deterministic and probabilistic deep learning models for inverse design of broadband acoustic cloak, Phys. Rev. Res. 3 (2021) 013142. https://doi.org/10.1103/PhysRevResearch.3.013142.

[31] H. Ding, X. Fang, B. Jia, N. Wang, Q. Cheng, Y. Li, Deep Learning Enables Accurate Sound Redistribution via Nonlocal Metasurfaces, Phys. Rev. Appl. 16 (2021) 064035. https://doi.org/10.1103/PhysRevApplied.16.064035.

[32] Z. Zeng, M. Zhang, C. Li, L. Ren, P. Wang, li jiawei_hust, D.W. Yang, Y. Pan, Simulation study on characteristics of acoustic metamaterials based on Mie and Helmholtz resonance for low-frequency acoustic wave control, J. Phys. D: Appl. Phys. (2021). https://doi.org/10.1088/1361-6463/ac0ad1.

[33] I. Davis, A. McKay, G.J. Bennett, A graph-theory approach to optimisation of an acoustic absorber targeting a specific noise spectrum that approaches the causal optimum minimum depth, J. Sound Vib. 505 (2021) 116135. https://doi.org/10.1016/j.jsv.2021.116135.

[34] J. Esteves, L. Rufer, D. Ekeom, S. Basrour, Lumped-parameters equivalent circuit for condenser microphones modeling, J. Acoust. Soc. Am. 142 (2017) 2121–2132. https://doi.org/10.1121/1.5006905.

[35] Y. Cheng, J.Y. Xu, X.J. Liu, One-dimensional structured ultrasonic metamaterials with simultaneously negative dynamic density and modulus, Phys. Rev. B 77 (2008) 045134. https://doi.org/10.1103/PhysRevB.77.045134

[36] N. Fang, D. Xi, J. Xu, M. Ambati, W. Srituravanich, C. Sun, X. Zhang, Ultrasonic metamaterials with negative modulus, Nat. Mater. 5 (2006) 452–456. https://doi.org/10.1038/nmat1644

[37] Y. Li, C. Shen, Y. Xie, J. Li, W. Wang, S.A. Cummer, Y. Jing, Tunable asymmetric transmission via lossy acoustic metasurfaces, Phys. Rev. Lett. 119 (2017) 035501. https://doi.org/10.1103/PhysRevLett.119.035501

[38] N. Jiménez, T.J. Cox, V. Romero-García, J.-P. Groby, Metadiffusers: Deep-subwavelength sound diffusers, Sci. Rep. 7 (2017) 5389. https://doi.org/10.1038/s41598-017-05710-5

[39] X. Yang, J. Yin, G. Yu, L. Peng, N. Wang, Acoustic superlens using Helmholtz-

resonator-based metamaterials, Appl. Phys. Lett. 107 (2015) 193505. https://doi.org/10.1063/1.4935589.

[40] J. Xia, X. Zhang, H. Sun, S. Yuan, J. Qian, Y. Ge, Broadband tunable acoustic asymmetric focusing lens from dual-layer metasurfaces, Phys. Rev. Appl. 10 (2018) 014016. https://doi.org/10.1103/PhysRevApplied.10.014016

[41] Y. Liu, W. Zhang, G. Cao, G. Zuo, C. Liu, F. Ma, Ultra-thin ventilated metasurface pipeline coating for broadband noise reduction☆, Thin-Walled Struct. 200 (2024) 111916. https://doi.org/10.1016/j.tws.2024.111916.

[42] H. Long, Y. Cheng, X. Liu, Reconfigurable sound anomalous absorptions in transparent waveguide with modularized multi-order Helmholtz resonator, Sci. Rep. 8 (2018) 1–9. https://doi.org/10.1038/s41598-018-34117-z

[43] Y. Yu, H. Jia, Y. Yang, H. Zhao, Q. Shi, P. Kong, J. Yang, K. Deng, Multi-order resonators for acoustic multiband asymmetric absorption and reflection, J. Appl. Phys. 131 (2022) 135102. https://doi.org/10.1063/5.0084450.

[44] X. Sun, H. Jia, Y. Yang, J. Yang, Low-frequency Broadband Sound Insulation Device Design Method Based on Deep Learning, J. Appl. Acoust. 42 (2023) 611–619.

[45] Q. Shi, Y. Yang, Z. Mei, Y. Lin, X. Li, P. Tian, P. Kong, H. Jia, J. Yang, K. Deng, Design and demonstration of composite mufflers based on dissipative and reactive units, Eng. Res. Express 5 (2023) 045029. https://doi.org/10.1088/2631-8695/ad03ae.

[46] C.R. Liu, J.H. Wu, X. Chen, F. Ma, A thin low-frequency broadband metasurface with multi-order sound absorption, J. Phys. D: Appl. Phys. 52 (2019) 105302. https://doi.org/10.1088/1361-6463/aafaa3.

[47] A.D. Pierce, Acoustics: An Introduction to Its Physical Principles and Applications, Springer International Publishing, Cham, 2019. https://doi.org/10.1007/978-3-030-11214-1.

[48] D.P. Kingma, J. Ba, Adam: A method for stochastic optimization, ArXiv preprint, ArXiv14126980 (2014). https://doi.org/10.48550/arXiv.1412.6980

[49] S. Ioffe, C. Szegedy, Batch normalization: Accelerating deep network training by reducing internal covariate shift, ArXiv preprint. ArXiv150203167 (2015). https://doi.org/10.48550/arXiv.1502.03167

[50] N. Srivastava, G. Hinton, A. Krizhevsky, I. Sutskever, R. Salakhutdinov, Dropout: a simple way to prevent neural networks from overfitting, J. Mach. Learn. Res. 15 (2014) 1929–1958.

[51] X. Sun, Y. Yang, H. Jia, J. Yang, Physics-aware training for the physical machine learning model building, The Innovation 3 (2022). https://doi.org/10.1016/j.xinn.2022.100287.

[52] K. Pearson, LIII. On lines and planes of closest fit to systems of points in space, Lond. Edinb. Dublin Philos. Mag. J. Sci. 2 (1901) 559–572. https://doi.org/10.1080/14786440109462720

[53] M. Yang, S. Chen, C. Fu, P. Sheng, Optimal sound-absorbing structures, Mater Horiz 4 (2017) 673–680. https://doi.org/10.1039/C7MH00129K.

[54] K.S. Tang, K.F. Man, S. Kwong, Q. He, Genetic algorithms and their applications, IEEE Signal Process. Mag. 13 (1996) 22–37. https://doi.org/10.1109/79.543973.

[55] T. Wang, S. Li, S.R. Nutt, Optimal design of acoustical sandwich panels with a genetic algorithm, Appl. Acoust. 70 (2009) 416–425. https://doi.org/10.1016/j.apacoust.2008.06.003.

[56] J. Kennedy, R. Eberhart, Particle swarm optimization, in: Proc. ICNN95 - Int. Conf. Neural Netw., IEEE, 1995, pp. 1942–1948 vol.4. https://doi.org/10.1109/ICNN.1995.488968.

[57] Jazzbin et al., geatpy: The genetic and evolutionary algorithm toolbox with high performance in Python, GitHub repository, 2020. https://github.com/geatpy-dev/geatpy.

[58] D. Liu, Y. Tan, E. Khoram, Z. Yu, Training Deep Neural Networks for the Inverse Design of Nanophotonic Structures, ACS Photonics 5 (2018) 1365–1369. https://doi.org/10.1021/acsphotonics.7b01377.

[59] Y. Yang, B. Yang, Equivalent circuit method based on complete magneto-mechanical coupling magnetostriction parameters for fixed magnetoelectric composites, Int. J. Mech. Sci. 199 (2021) 106411. https://doi.org/10.1016/j.ijmecsci.2021.106411.

[60] A. Dorodnyy, S.M. Koepfli, A. Lochbaum, J. Leuthold, Design of CMOS-compatible metal–insulator–metal metasurfaces via extended equivalent-circuit analysis, Sci. Rep. 10 (2020) 17941. https://doi.org/10.1038/s41598-020-74849-5.

[61] A. Khajevandi, H. Oraizi, Utilizing interdigital and supershape geometries for the design of frequency selective surfaces with high angular and polarization stabilities, Sci. Rep. 12 (2022) 7054. https://doi.org/10.1038/s41598-022-10960-z.